\documentclass[conference]{IEEEtran}
\IEEEoverridecommandlockouts

\usepackage{cite}
\usepackage[numbers]{natbib}
\usepackage{amsmath,amssymb,amsfonts}
\usepackage{algorithmic}
\usepackage{graphicx}
\usepackage{textcomp}
\usepackage{todonotes}
\usepackage{tabularx}
\usepackage{float}
\usepackage{balance}
\usepackage{pifont}
\usepackage[inline]{enumitem}
\usepackage{orcidlink}
\usepackage{framed}
\usepackage{booktabs}
\usepackage{caption}
\usepackage{subcaption}

\usepackage{tcolorbox}
\usepackage{xcolor}
\tcbuselibrary{skins, breakable}

\definecolor{rqcolor}{RGB}{83, 74, 183} 
\newenvironment{rqbox}[3]{%
  \begin{tcolorbox}[
    enhanced,
    breakable,
    colframe=#1,
    colback=#1!6!white,
    colbacktitle=#1,
    coltitle=white,
    fonttitle=\bfseries\small,
    title={#2\quad #3},
    left=4pt, right=4pt, top=2pt, bottom=2pt,
    boxrule=0.5pt,
    arc=3pt,
    toptitle=2pt,
    bottomtitle=2pt,
  ]
}{%
  \end{tcolorbox}
}

\newcommand{\subrq}[3]{%
  \vspace{2pt}
  \begin{tcolorbox}[
    enhanced,
    colframe=#1!40!white,
    colback=white,
    left=4pt, right=4pt, top=2pt, bottom=2pt,
    boxrule=0.4pt,
    leftrule=2.5pt,
    arc=2pt,
    before skip=1pt,
    after skip=1pt,
  ]
  \small
  \noindent
  \makebox[3.5em][l]{\textcolor{#1}{\textbf{#2}}}%
  \parbox[t]{\dimexpr\linewidth-3.5em\relax}{#3}
  \end{tcolorbox}%
}

\def\BibTeX{{\rm B\kern-.05em{\sc i\kern-.025em b}\kern-.08em
    T\kern-.1667em\lower.7ex\hbox{E}\kern-.125emX}}

\definecolor{mairieli}{HTML}{ff75ad}

\definecolor{pien}{HTML}{c44fe8}

\definecolor{christoph}{HTML}{4fe866}

\begin{document}


\title{How Humans, Bots, and Agents Communicate About Vulnerabilities in Pull Requests}

\author{\IEEEauthorblockN{Pien Rooijendijk}
\IEEEauthorblockA{
\textit{Radboud University}\\
The Netherlands}
\and
\IEEEauthorblockN{Christoph Treude}
\IEEEauthorblockA{
\textit{Singapore Management University}\\
Singapore}
\and
\IEEEauthorblockN{Mairieli Wessel}
\IEEEauthorblockA{
\textit{Radboud University}\\
The Netherlands}
}

\maketitle

\begin{abstract}
Developers may reference vulnerabilities in pull request discussions through both explicit identifiers, such as CVEs or GHSAs, and implicit security-related language (e.g., ``unauthorized access'' or ``SQL injection''). Prior work has primarily focused on explicit identifiers, potentially overlooking vulnerability discussions that lack formal references. Bots and coding agents are becoming more common in pull requests, raising new questions about how different accounts communicate about vulnerabilities. In this registered report, we describe our planned study of vulnerability communication in pull requests by humans, bots, and coding agents. Building on the AIDev-pop dataset, we analyze explicit vulnerability references and implicit security-related signals across pull request titles, descriptions, review comments, commit messages, and timeline discussions. We further investigate whether these references are associated with vulnerabilities introduced or fixed in the modified code and how they relate to pull request review activity and outcomes. This study contributes a large-scale empirical investigation of vulnerability communication practices in modern software development.
\end{abstract}


\section{Introduction}

Software vulnerabilities are routinely identified, cataloged, and communicated through standardized identifier systems such as Common Vulnerabilities and Exposures (CVE), Common Weakness Enumeration (CWE), GitHub Security Advisories (GHSA), and ecosystem-specific databases. These identifiers support coordination across projects, tools, and communities by providing shared references for discussing weaknesses, tracking disclosures, and linking fixes to known issues. Prior work has shown that such identifiers are frequently referenced in software repository artifacts, including commits, issues, and pull requests~\cite{nakano2020quantitative,bhandari2021,hommersom2024}.

However, developers do not communicate about vulnerabilities only through formal identifiers. Security concerns may also be discussed through implicit language, such as ``unauthorized access,'' ``SQL injection,'' or ``insecure behavior,'' without referencing a CVE or GHSA. Prior work suggests that vulnerability discussions often occur before formal disclosure and that communication practices vary across repositories and development contexts~\cite{liu2025empirical,ayala2025mixed,reis2023security,croft2022empirical}.

Previous work has primarily used vulnerability identifiers to study security-related development activity, including vulnerability-fixing commits, dependency management, disclosure practices, and vulnerability response times~\cite{antal2020exploring,kumar2024comprehensive,kancharoendee2025categorizing}. Other studies explored automated detection of security-related discussions using keyword-based and machine learning approaches~\cite{zhou2017automated,cipollone2025automating}. However, these studies focus on identifying security-related artifacts, such as GitHub issues, rather than understanding how vulnerabilities are communicated within pull request discussions.

Bots and coding agents are becoming increasingly common in pull request workflows. Recent work showed that coding agents can generate security-related pull requests and that bots frequently reference vulnerability identifiers in automated dependency updates and security patches~\cite{rooijendijk2026said,siddiq2026security,wang2025automated}. While our prior work examined how humans, bots, and coding agents use explicit vulnerability identifiers~\cite{rooijendijk2026said}, it did not consider implicit security-related signals or whether such references correspond to actual vulnerabilities in the modified code. As a result, little is known about how different actors communicate about vulnerabilities beyond formal identifiers, whether these references are associated with vulnerabilities being introduced or fixed, and how they influence pull request review and outcomes.

In this registered report, we describe our plans for investigating how vulnerabilities are referenced in GitHub pull requests by humans, bots, and coding agents. We build on the AIDev-pop dataset~\cite{li2025rise} and extend it with additional pull requests collected from the same repositories. We analyze explicit vulnerability references and implicit security-related signals across pull request titles, descriptions, review comments, commit messages, and timeline discussions. Explicit references are identified using regular expression matching for standardized vulnerability identifiers, while implicit signals are detected using a validated keyword-based approach for security-related language~\cite{zhou2017automated}.

This study contributes a large-scale analysis of explicit vulnerability references and implicit security-related signals in pull requests across humans, bots, and coding agents. In addition, we provide a dataset and analysis pipeline for studying vulnerability communication in pull request discussions.

\section{Background \& Related Work}
\begin{figure*}[t]
    \centering

    \begin{subfigure}[b]{0.48\textwidth}
        \centering
        \includegraphics[width=\textwidth]{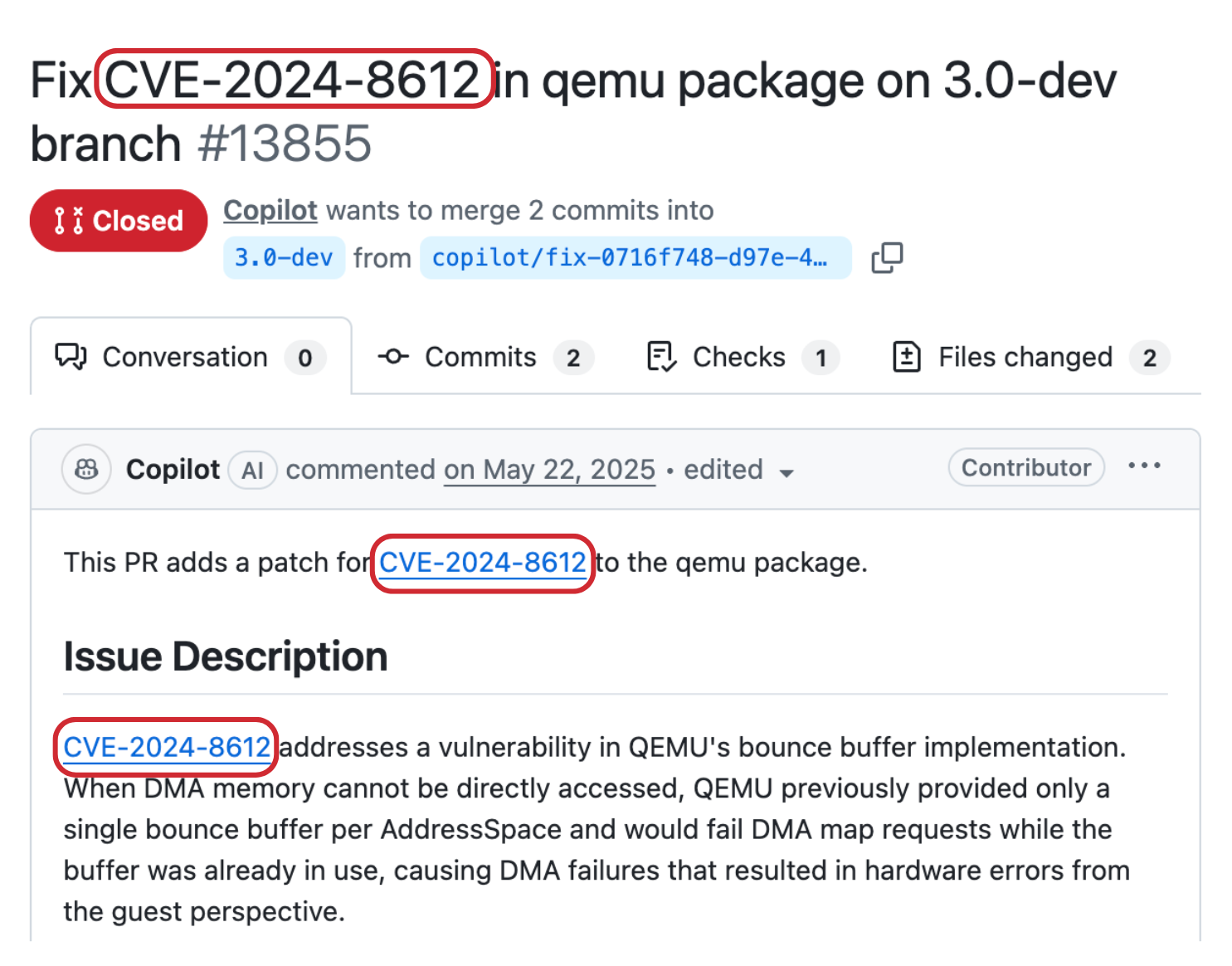}
        \caption{Explicit vulnerability references}
        \label{fig:explicit}
    \end{subfigure}
    \hfill
    \begin{subfigure}[b]{0.48\textwidth}
        \centering
        \includegraphics[width=\textwidth]{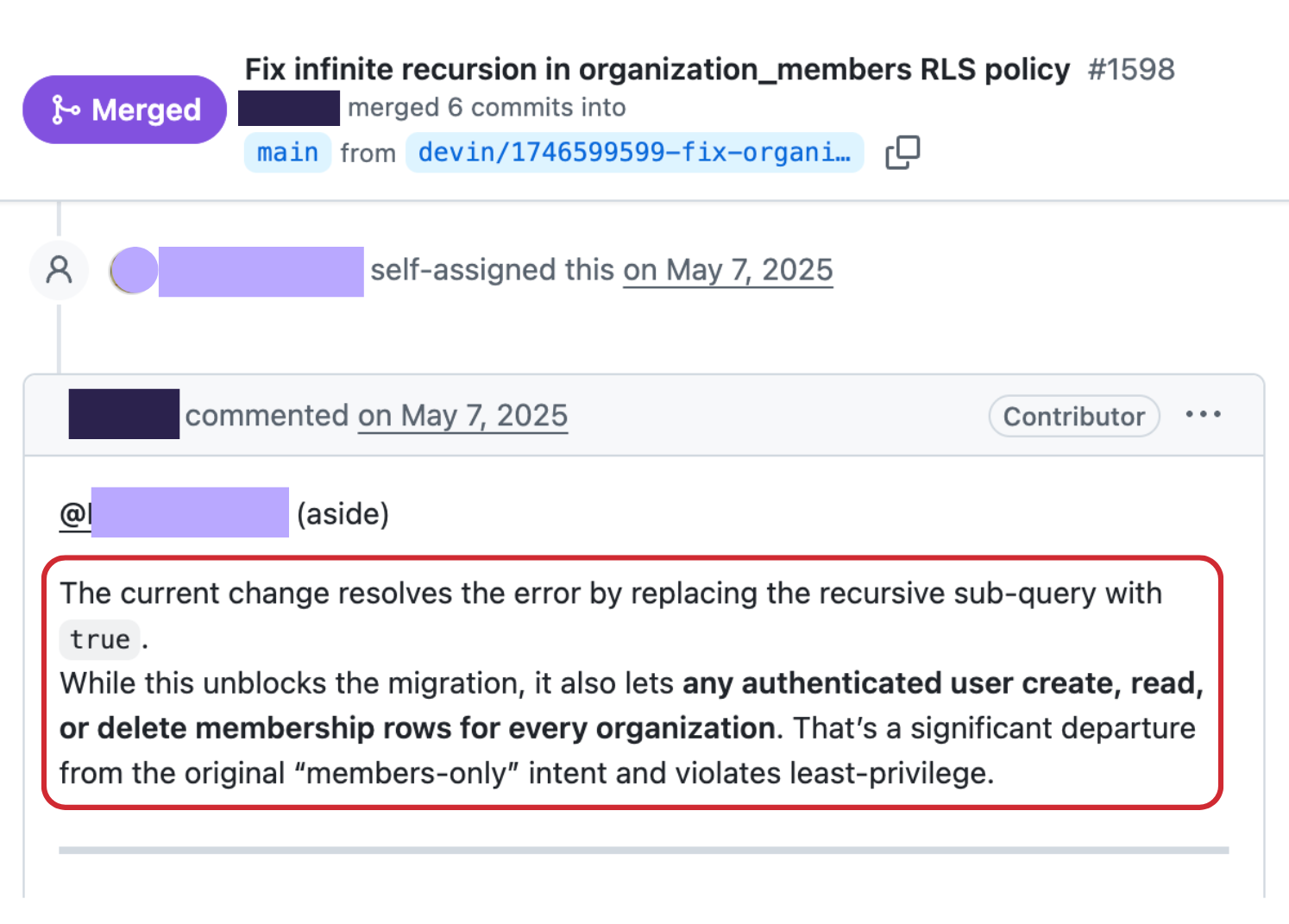}
        \caption{Implicit security-related signals}
        \label{fig:implicit}
    \end{subfigure}
\caption{Examples of \textbf{explicit} and \textbf{implicit} references \& signals in pull requests.}
\end{figure*}

\subsection{Vulnerability Signals in Software Repositories}

Software vulnerabilities are frequently discussed in pull requests (PRs) through both \textbf{explicit vulnerability references}, which directly mention standardized identifiers such as CVEs, CWEs, GHSAs, or RUSTSEC entries, and \textbf{implicit security-related signals}, which describe security concerns without referencing a formal identifier. Explicit references provide structured and verifiable links to vulnerability databases and advisory systems and have been widely used to study vulnerability fixes and security-related development activity. Figure~\ref{fig:explicit} shows an example of an explicit vulnerability reference in a pull request by Copilot.\footnote{https://github.com/microsoft/azurelinux/pull/13855}

Not all security-related discussions include explicit identifiers. Developers may instead describe vulnerabilities through natural language, for example by referring to insecure behaviour, attack scenarios, or mitigation strategies without mentioning a known CVE, CWE, or GHSA entry~\cite{reis2023security,croft2022empirical}. We refer to these as \textbf{implicit security-related signals}. Figure~\ref{fig:implicit} presents an example where a reviewer highlights a potential authorization issue without referencing a formal vulnerability identifier.\footnote{https://github.com/liam-hq/liam/pull/1598}

Prior work has investigated how vulnerabilities are referenced and discussed in software repositories. Many studies rely on explicit vulnerability identifiers such as CVEs, CWEs, and GHSAs to trace vulnerability fixes, disclosure practices, dependency management, and other security-related development activities~\cite{nakano2020quantitative,bhandari2021,hommersom2024}. Recent work also showed that vulnerability discussions frequently occur in repository artifacts such as issues, commits, and pull requests, sometimes before formal disclosure~\cite{liu2025empirical,ayala2025mixed}.

While most of this work focuses on explicit vulnerability identifiers, other studies have explored how security-related content can be identified from natural-language discussions. Zhou and Sharma~\cite{zhou2017automated} proposed a keyword-based approach for identifying security-related discussions from commit messages and bug reports, while Cipollone et al.~\cite{cipollone2025automating} used Transformer-based models to identify vulnerability-related GitHub issues. Together, these studies suggest that analyses relying solely on explicit identifiers may overlook security-related discussions that lack formal vulnerability references.

\subsection{Security Related Pull Requests}

Research on security-related pull requests has largely focused on dependency remediation, vulnerability classification, and review outcomes. Studies of dependency management bots found that automated security pull requests are widely used but are not always merged or acted upon by developers~\cite{alfadel2021,mohayeji2023}. More recently, work on coding agents showed that agents increasingly participate in security-related development activities and generate pull requests involving vulnerability-related changes~\cite{siddiq2026security,wang2025automated}.

Our prior work~\cite{rooijendijk2026said} examined how humans, bots, and coding agents use explicit vulnerability identifiers in pull requests. In contrast, the present study investigates both explicit and implicit vulnerability references, examines their relationship with vulnerabilities detected in the modified code, and analyzes their association with pull request review activity and outcomes.

\section{Research Questions}

We investigate how explicit and implicit vulnerability references are communicated in pull requests by humans, bots, and coding agents through the following research questions:

\begin{rqbox}{rqcolor}{RQ1}{How are vulnerability references distributed in pull requests in terms of implicit and explicit references?}
  \subrq{rqcolor}{RQ1a}{How frequently do implicit and explicit vulnerability references occur in pull requests?}
  \subrq{rqcolor}{RQ1b}{How does the distribution of vulnerability references vary across actor types (humans, bots, agents)?}
  \subrq{rqcolor}{RQ1c}{How does the distribution of vulnerability references vary across pull request components?}
\end{rqbox}

Vulnerabilities may be discussed both through explicit identifiers and implicit natural-language descriptions. However, little is known about how these references are distributed across pull request artifacts and actor types. Understanding these distributions is important because repository mining studies and security tooling often rely primarily on explicit identifiers, potentially overlooking vulnerabilities that are discussed only implicitly. To answer this question, we identify explicit references using regular expression patterns and implicit references using a validated keyword-based approach applied to pull request artifacts.


\begin{tcolorbox}[
  colback=rqcolor,
  colframe=rqcolor,
  coltext=white,
  fontupper=\bfseries\small,
  boxrule=0.4pt,
  arc=2pt,
  left=3pt,
  right=3pt,
  top=3pt,
  bottom=3pt
]
RQ2\hspace{0.8em}How are implicit and explicit vulnerability references associated with static-analysis-detectable security changes of vulnerability detection, as reflected by indicators of actual vulnerabilities in code changes?
\end{tcolorbox}

Recent studies reported that coding agents frequently generate security-related pull requests, but also produce false positives and security-related claims that do not correspond to actual vulnerabilities~\cite{steenhoek2025closing}. This raises the question of whether the way vulnerabilities are referenced in pull requests is associated with the presence of actual security-relevant code changes. We therefore compare vulnerability references against static analysis results from the code before and after the pull request changes.

\begin{rqbox}{rqcolor}{RQ3}{How are implicit and explicit vulnerability references associated with the social response to pull requests?}
\subrq{rqcolor}{RQ3a}{How are implicit and explicit vulnerability references associated with pull request review activity?}
\subrq{rqcolor}{RQ3b}{How are implicit and explicit vulnerability references associated with pull request outcomes?}
\end{rqbox}

The way vulnerabilities are communicated may influence how pull requests are reviewed and evaluated. Explicit identifiers provide direct links to external advisories and vulnerability databases, while implicit references require reviewers to interpret and validate security-related claims from the discussion context. To investigate whether these differences affect pull request evaluation, we analyze review activity, discussion patterns, response times, and pull request outcomes across different vulnerability reference types and actor categories.

\section{Dataset}

Our study builds on the AIDev-pop dataset, a curated subset of the AIDev dataset introduced by Li et al.~\cite{li2025rise}. AIDev captures pull requests (PRs) authored by autonomous coding agents on GitHub, including OpenAI Codex, Devin, GitHub Copilot, Cursor, and Claude Code. The original dataset contains more than 456k agent-authored PRs across over 61k repositories and includes rich repository, review, timeline, and commit-level metadata, with data collected up to August 1, 2025.

We use AIDev-pop, which filters the dataset to include repositories with more than 500 GitHub stars, focusing on popular and active open-source projects. As shown in Table~\ref{tab:dataset}, AIDev-pop contains 33,078 pull requests across 2,807 repositories authored by five coding agents. AIDev-pop includes pull request titles and descriptions, review comments, commit messages, timeline events, and commit level metadata such as patches and file changes.

\begin{table}[h]
\centering
\caption{Distribution of pull requests and repos by agent.}
\label{tab:dataset}
\small
\begin{tabular}{lrr}
\toprule
Agent & Pull Requests & Repositories \\
\midrule
OpenAI Codex & 21,779 & 1,248 \\
Devin & 4,827 & 288  \\
GitHub Copilot & 4,970 & 1,012 \\
Cursor & 1,541 & 327  \\
Claude Code & 459 & 213  \\
\midrule
Total & 33,078 & 2,807  \\
\bottomrule
\end{tabular}
\end{table}

To account for vulnerability references beyond agent-authored PRs, we collect all pull requests created during the same observation period as AIDev-pop across all repositories in the dataset. This extension allows us to compare vulnerability references across humans, bots, and coding agents within the same repository ecosystems.

\section{Execution Plan}

\begin{figure*}[t]
    \centering
    \includegraphics[width=\textwidth]{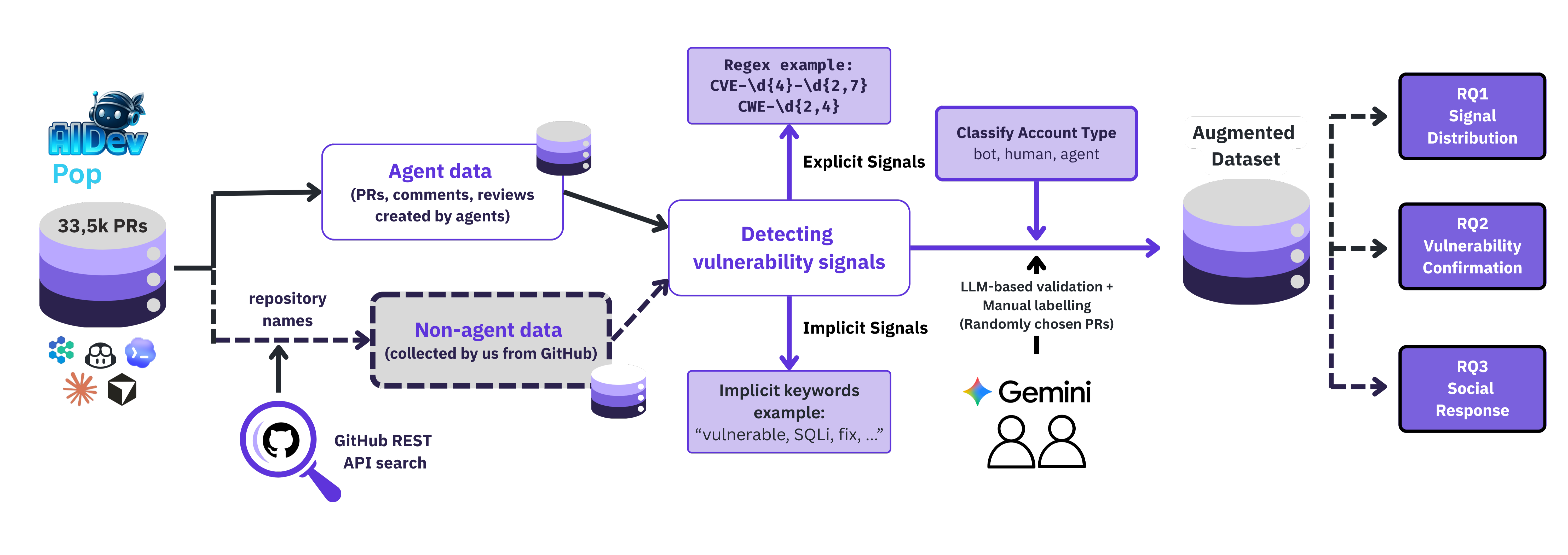}
    \caption{Our study's execution plan}
    \label{fig:execution_plan}
\end{figure*}

Figure~\ref{fig:execution_plan} provides an overview of the study design.

\subsection{Vulnerability Signals Detection}
To identify vulnerability references, we analyze pull request artifacts at a fine-grained level rather than classifying pull requests as a whole. Specifically, we analyze pull request titles, descriptions, review comments, commit messages, and timeline messages individually. Each detected reference is associated with both the artifact in which it appears and the account responsible for authoring that artifact. This enables us to study how vulnerability references are communicated across different interaction contexts and account types.

\subsubsection{Explicit References}
\begin{table}[ht]
    \centering
    \caption{Regex patterns used for explicit vulnerability signal detection.}
    \label{tab:regex}
    \renewcommand{\arraystretch}{1.1}
    \scriptsize
    \begin{tabular}{@{}lp{3.3cm}l@{}}
        \hline
        \textbf{ID} & \textbf{Regex pattern} & \textbf{Source} \\
        \hline
        CVE & CVE$-\backslash d\{4\}-\backslash d\{2,7\}$ & cve.mitre.org \\
        CWE & CWE$-\backslash d\{2,4\}$ & cwe.mitre.org \\
        GHSA & GHSA$-[a-z0-9]\{4\}-[a-z0-9]\{4\}-[a-z0-9]\{4\}$ & github.com/advisories \\
        GO & GO$-\backslash d\{4\}-\backslash d\{2,4\}$ & vuln.go.dev \\
        RUSTSEC & RUSTSEC$-\backslash d\{4\}-\backslash d\{4,7\}$ & rustsec.org \\
        OSV & OSV$-\backslash d\{4\}-\backslash d\{4,7\}$ & osv.dev/list \\
        MAL & MAL$-\backslash d\{4\}-\backslash d\{4,7\}$ & github.com/ossf/malicious-packages \\
        USN & USN$-\backslash d\{4\}-\backslash d\{1,2\}$ & ubuntu.com/security/cves \\
        \hline
    \end{tabular}
\end{table}

Explicit vulnerability references are comparatively easier to identify because they follow standardized textual patterns. We therefore detect explicit references using regular expression matching. We use a keyword-based approach to identify candidate security-related signals, which are subsequently validated through LLM-assisted annotation and manual review.

We apply the regular expression patterns shown in Table \ref{tab:regex} independently to all analyzed pull request artifacts. In this study, we rely on the Open Source Vulnerability (OSV) specification\footnote{https://ossf.github.io/osv-schema}, which defines a common interchange format for vulnerabilities and supports identifiers from multiple vulnerability databases and ecosystems. The AIDev-pop dataset references eight OSV-supported identifier schemes, consistent with prior work~\cite{rooijendijk2026said}.

Each detected explicit reference is stored with the matched identifier, the pull request artifact where it was found, the associated pull request, the surrounding text, the author's account classification, and the GitHub URL for verification. When multiple identifiers appear within the same artifact, each identifier is recorded separately.

\subsubsection{Implicit Signals}

Unlike explicit references, implicit vulnerability signals do not follow fixed identifier patterns and instead rely on natural-language descriptions of security-related concerns. 
To detect them, we adopt the keyword-based methodology proposed by Zhou and Sharma~\cite{zhou2017automated}. The approach combines security-related keywords with contextual filtering to distinguish security discussions from generic bug-fixing language.

We apply this approach to the same pull request artifacts analyzed for explicit references. Each detected implicit reference is stored with the same metadata collected for explicit references, including the artifact type, associated pull request, surrounding text, actor classification, and verification URL. Pull requests may contain both explicit and implicit references across different artifacts.

\subsection{Manual Evaluation}
Following Rabbi et al.’s~\cite{rabbi2026insights} LLM-based validation method, we validate the found explicit and implicit vulnerability signals using the Gemini-2.0-flash model. The model outputs a binary label, yes or no, if the signal references or communicates about a vulnerability. If Gemini-2.0-flash is no longer available at the time of data collection, we will use the closest available successor model and report this change in the final manuscript.
To assess the reliability of the LLM annotations, we draw two representative samples of 360 pull requests each, one containing explicit signals and one containing implicit signals. This number is based on the 5,465 pull requests containing explicit security signals identified in our previous work. This sample size is determined under the assumption of a 95\% confidence level and a 5\% margin of error. The samples are stratified across actor types and signal categories. Two authors independently code this subset as having a vulnerability signal or not, based on the complete pull request contents. The inter-rater reliability is measured by Cohen's $\kappa$~\cite{vieira2010cohen}, which indicates the agreement level between the human annotators. Cohen's $\kappa$ is computed before disagreement resolution. Disagreements will be discussed to agree on a shared coding scheme. The resulting human annotations serve as ground truth and are compared against the LLM-generated labels to evaluate the reliability of the automated annotations using accuracy, precision, recall, and F1-score \cite{rao2026agreement}.  The results are reported alongside the main findings.

\subsection{Account Type Classification}


We classify the author of each detected vulnerability reference as a human, bot, or coding agent.

\noindent
\textbf{\textit{Coding agent accounts:}}
To identify coding agent accounts, we use the user accounts associated with autonomous coding agents (e.g., \texttt{devin-ai-integration[bot]}), which were annotated in the original AIDev-pop dataset by Li et al.~\cite{li2025rise}. References appearing in artifacts authored by these accounts were classified as agent-generated references.

\noindent
\textbf{\textit{Bot accounts:}}
Among the remaining accounts, we identified bots using three high-precision signals commonly used in prior work~\cite{bothunter,chidambaram2023dataset,rooijendijk2026said}. First, we identified accounts explicitly tagged by GitHub as bots, such as \texttt{dependabot[bot]}. Second, we identified usernames containing the string ``bot.'' Third, we matched accounts against a curated list of 385 well-known bot accounts reported by Chidambaram et al.~\cite{chidambaram2023dataset}. We additionally performed manual verification to improve the reliability of the classification, where we classified a representative random sample of 242 accounts. This is based on the unique 654 bot accounts found in our previous work, using a 95\% confidence level and a 5\% margin of error. 

\noindent
\textbf{\textit{Human accounts:}}
All remaining accounts that are not classified as coding agents or bots are then classified as human contributors.

\subsection{Data Analysis}

The analyses are organized around the three research questions (RQ1–RQ3). Comparisons across actor types, pull request components, and other categories provide additional context for interpreting the results. Given the exploratory nature of this study, subgroup analyses are interpreted with caution and reported together with effect sizes.

\subsubsection{RQ1: Distribution of Vulnerability References}
In \textbf{$\mathbf{RQ_1}$}, we compute the distribution of explicit and implicit vulnerability references across all PRs in AIDev-pop, by PR component. We report the proportion of security-mentioning PRs that use only implicit language, only explicit identifiers, or both. To compare the distribution of vulnerability references across actor types and pull request components, we use chi-square tests of independence and report Cramér's V as an effect size. For repository-level analyses, we additionally report results aggregated by repository.

\subsubsection{RQ2: Quality of Vulnerability Detection}
To answer \textbf{$\mathbf{RQ_2}$} we use the static analysis and software composition analysis (SCA) tool of Semgrep\footnote{https://semgrep.dev} to evaluate the original code and the changed code to assess whether the proposed change introduces or fixes a vulnerability. We configure Semgrep with the default rule sets and the four severity levels: \textit{critical, high, medium} and \textit{low}. We selected Semgrep because it supports the programming languages represented in AIDev-pop and enables reproducible vulnerability scans~\cite{kuszczynski2023comparative,bennett2024semgrep}.

For each PR containing a vulnerability signal, we perform a static analysis scan on all commits in the PR. For pull requests with multiple commits, we also scan intermediate commits to trace how vulnerabilities evolve throughout the pull request. However, the primary classification of vulnerabilities as introduced, fixed, or unchanged is based on comparisons between the pull request's base and head commits. The \textit{before} version is defined as the base commit of the PR, the commit that is at the head of the branch at the time the PR was opened, which is the state of the code before any changes were made. To track vulnerabilities across commits, findings are matched using the same vulnerability ID and relative file path to the vulnerable code. If a vulnerability exists in the \textit{before} version of the pull request but was resolved in one of the following commits within the modified lines of the PR diff, it is classified as \textit{fixed}. If a vulnerability appears in the head commit but was absent in the base commit, it is classified as \textit{introduced}. When multiple findings are present within a PR, we aggregate the results at the PR level based on whether the vulnerabilities were introduced, fixed, or unchanged. The \textit{after} version is defined as the head commit of the PR, meaning the last commit of the PR at the time of closure, regardless of whether the PR was merged. For merged PRs, this means the code that was integrated into the branch. For closed but unmerged PRs, the head commit has the final changes and is scanned to assess whether the vulnerability signal corresponded to a vulnerability in the changed code, even if those changes were never accepted.

This analysis evaluates how accurately the presence of an implicit or explicit vulnerability signal in PR descriptions reflects its reliability. A PR might contain a detailed security explanation without actually addressing a vulnerability, whereas another may describe a concrete vulnerability without referencing a formal identifier. By linking these outcomes to the type of vulnerability signal used in the PR, we can analyze whether explicit or implicit references are more strongly associated with actual vulnerability fixes, and whether this relationship varies across actor types. To assess these relationships, we fit regression models using the Semgrep outcome (introduced, fixed, or unchanged) as the dependent variable and vulnerability reference type and actor type as independent variables. The models additionally control for pull request characteristics, including the number of changed files, lines added or deleted, and the number of commits. We account for repository-level clustering to reduce the influence of repository-specific development practices.

\subsubsection{RQ3: Social Response}

To answer \textbf{$\mathbf{RQ_3}$}, we compare review activity and pull request outcomes across different vulnerability reference types and account categories. Specifically, we analyze the number of comments, number of reviewers, pull request status (merged, closed, or open), and time until the first response for pull requests containing explicit and implicit vulnerability references. We consider the final state of each pull request at the cut-off date of August 1, 2025.

We further investigate whether implicit references are associated with longer discussions, delayed responses, or lower merge rates compared to explicit references. Since review activity may also depend on pull request characteristics unrelated to vulnerability communication, we control for factors such as the number of changed files, the number of lines added or deleted, the repository, and the programming language. This helps account for differences caused by pull request size, complexity, or repository context. We use statistical models appropriate to each outcome variable. Pull request status is analyzed using logistic regression, review activity measures, such as the number of comments and reviewers, are analyzed using count-based regression models, and time until first response is analyzed using time-to-event models. All models include the control variables described above and account for repository-level clustering.

\section{Threats to Validity}

\textbf{Internal Validity.}
Our identification of explicit vulnerability references relies on regular expression matching. Although identifiers such as CVEs, CWEs, GHSAs, and Go vulnerability IDs follow standardized formats, malformed identifiers may go undetected.
Detecting implicit vulnerability references is more challenging because it relies on natural-language descriptions rather than fixed patterns. Although we adopt the validated keyword-based approach of Zhou and Sharma~\cite{zhou2017automated}, implicit references may still produce false positives or false negatives. To mitigate this threat, detected signals are validated through LLM-assisted annotation and a manually reviewed sample of pull requests across different account types and reference categories.
Account classification may also introduce inaccuracies. Coding agent accounts rely on the AIDev-pop annotations of Li et al.~\cite{li2025rise}, while bot identification combines GitHub bot tags, username heuristics, and a curated list of known bot accounts. Consequently, some automated accounts may still be misclassified.
RQ2 further depends on Semgrep and Semgrep Supply Chain to assess whether pull request changes introduce, fix, or preserve vulnerabilities. Static analysis and software composition analysis tools may generate false positives or fail to detect vulnerabilities, particularly for complex or context-dependent security issues.

\textbf{External Validity.}
We focus on popular open-source GitHub repositories with more than 500 stars. As a result, our findings may not generalize to smaller repositories, private repositories, or software projects hosted outside GitHub.
The AIDev-pop dataset focuses on repositories that actively use coding agents and therefore contains a higher proportion of agent-authored pull requests than would be expected in the broader GitHub ecosystem. Consequently, our findings should be interpreted as characterizing vulnerability communication practices within repositories where coding agents are present rather than as prevalence estimates across all open-source projects. To mitigate this threat, we report observed actor distributions, rely on proportional measures rather than raw counts, and account for repository-level variation in the analyses.
The prevalence and style of vulnerability references may also vary across programming languages, ecosystems, and repository communities. Finally, our analysis focuses on pull request artifacts and does not capture vulnerability communication occurring through external channels such as mailing lists, private disclosures, chat platforms, or issue trackers outside GitHub.

\section{Conclusion}

This registered report presents a study of explicit vulnerability references and implicit security-related signals in GitHub pull requests authored by humans, bots, and coding agents. We examine how these references are communicated, how they relate to vulnerabilities in the modified code, and how they are associated with review activity and pull request outcomes. Future work could investigate whether vulnerability references correspond to incorrect or unsupported security claims, particularly in AI-generated pull requests.

\section{Acknowledgements}
Pien's work is supported by the Dutch science foundation NWO through the KIC ``Find2Fix'' project (No. KICH1.VE05.23.008).
Special thanks to Fred \includegraphics[height=0.9em]{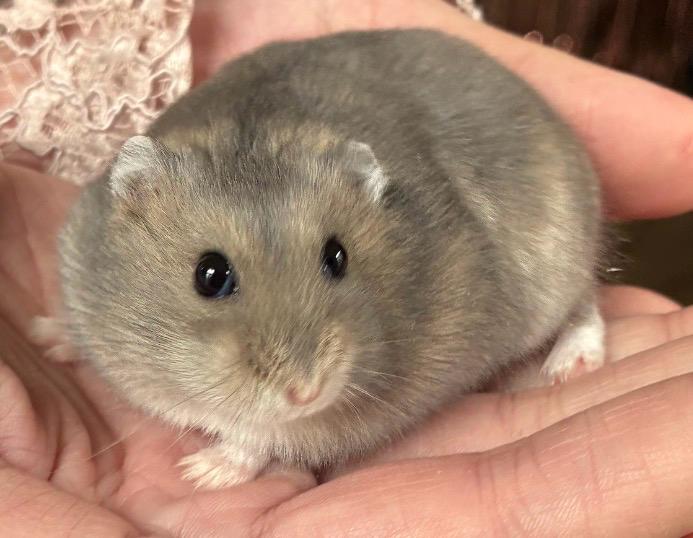} and Milo \includegraphics[height=1em]{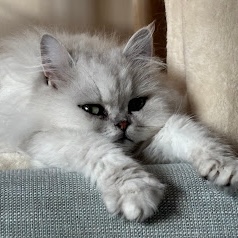}.

\balance
\bibliography{IEEEfull}

\end{document}